\documentclass[final,5p,times,twocolumn]{elsarticle}
\usepackage{lineno}
\usepackage{graphicx}
\biboptions{numbers,sort&compress} 
\usepackage[integrals]{wasysym} 
\modulolinenumbers[5]
\usepackage{natbib}
\usepackage{etoolbox}
\AtBeginEnvironment{thebibliography}{%
  \setlength{\parskip}{0pt}%
  \setlength{\itemsep}{0pt}%
}

\usepackage{booktabs,siunitx,threeparttable}
\usepackage{amssymb, hyperref}
\usepackage{amsmath}
\usepackage{lineno}

\begin{document}

\begin{frontmatter}

\title{Optical-field-induced dips and splits in nonlinear spectra of selective
       reflection from high-density atomic vapor}

\author[JIHT]{Vladimir Sautenkov}
\author[JIHT,HSE]{Sergey Saakyan\corref{cor}}
\ead{saakyan@ihed.ras.ru}
\author[JIHT]{Andrei Bobrov}
\author[JIHT]{Boris B. Zelener}

\address[JIHT]{Joint Institute for High Temperatures, Russian Academy of 
Sciences (JIHT RAS), Izhorskaya St. 13 Bld. 2, Moscow 125412, Russia}
\address[HSE]{National Research University Higher School of Economics (NRU HSE),
Myasnitskaya Ulitsa 20, Moscow 101000, Russia}
\cortext[cor]{Corresponding author}

\begin{abstract}
We discuss nonlinear spectra of selective reflection from high-density rubidium
atomic vapor, where the self-broadening of the resonant transition 
$5S_{1/2}-5P_{3/2}$ dominates over the Doppler width. 
In the experiments, the hole-burning technique with probe and pump lasers is used. 
The reflection of weak probe beam is investigated at four atomic densities in 
the range $(1.2\text{--}3.6)\times10^{17}$~cm$^{-3}$ and various pump beam
intensities. 
To enhance the spectral resolution, the frequency derivative 
$\text{d}R/\text{d}\nu$ of the reflection coefficient $R$ is analyzed.
Increasing the atomic number density changes the character of self-broadening
from inhomogeneous to homogeneous. 
At the highest density, the strong pump field splits the observed spectra into two
homogeneously broadened symmetric resonances. 
The appearance of the optical-field-induced resonances can be explained within the framework of "dressed atomic states" approach. 
At lower densities the spectral profiles are inhomogeneously
broadened. 
Spectral profiles of the frequency derivative are separated by optically 
saturated dips. 
The width of such dips is a combination of the homogeneous component of 
self-broadening and intensity-dependent field broadening.
Careful study of the transition from inhomogeneous to homogeneous broadening may
initiate further development of the theory of interatomic interactions in
high density atomic gas media.

\end{abstract}


\end{frontmatter}


\section{Introduction}
In high-density atomic gas, where dipole-dipole self-broadening of the resonance transitions is much larger than the Doppler width, the thermal motion of atoms can be neglected\cite{Lewis1980, BoydPRL1991}. 
The self-broadening $\Gamma_0$ in such gas of atomic density $N$ can be estimated by using a relation $\Gamma_0 = K N$, where factor $K\sim10^{-16}$~GHz\,cm$^3$.
The theory of self-broadening was extended in the framework of disordered excitons in Ref.~\cite{MukamelPRA1994}. 
The developed approach was applied to the study of dense two-level atomic gases. 
The main part of the Hamiltonian used described dipole-dipole interactions between the ground $S$-state and excited $P$-state atoms. 
Possible contributions of interactions between atoms in the same quantum states were neglected. 
It was noted that spectral profiles of resonance transitions were usually classified as either homogeneous or inhomogeneous. 
A spectral line shape was considered to be homogeneously broadened when the dominant broadening mechanism was due to the binary collisions of the moving atoms (collisional broadening), and inhomogeneous if the spectral width was mainly caused by a distribution of transition frequencies (static broadening). 
The inhomogeneous self-broadening induced by many-particle interactions was explained by the developed model of frequency shifted excitons with long-range interactions. 
The concept of the Lorentz local field in a disordered medium was also used. 
In the realm of linear optics, it is impossible to distinguish self-broadened line shapes in a dense atomic gas. 
In the discussed approach both broadening mechanisms gave Lorentzian-like spectral profiles and, moreover, the calculated static and collision widths are of comparable magnitudes, with a fixed ratio of 3:8, regardless of the atomic density. 

Notably, there must be a threshold density above which the binary collisions become interrupted by third particles. 
Such an event would occur when the average interparticle distance becomes of less or the same order of magnitude as the Weisskopf radius.
A density $N_\mathrm{W}$, which defined by Weisskopf radius for potassium atomic vapor, was estimated to be $10^{18}$~cm$^{-3}$. 
At densities $N\ge N_\mathrm{W}$, the dipole-dipole interactions are altered. 
The discussed theoretical treatment~\cite{MukamelPRA1994} was performed for much less densities, $N\ll N_\mathrm{W}$. 
Under these conditions the short-range deviations from the dipole-dipole interactions (influence of higher multipoles) were neglected. 
Predicted homogeneous or inhomogeneous character of the spectral profile of atomic transition could be tested by nonlinear laser spectroscopy methods such as photon echoes or hole burning~\cite{MukamelJCP1991}. 
The publications~\cite{MukamelPRA1994, MukamelJCP1991} motivated authors of the Ref.~\cite{VASPRL1996} to perform an experimental study of a nonlinear selective reflection from the dense rubidium atomic vapor.

In the experiment on rubidium vapor~\cite{VASPRL1996} the probe and pump tunable lasers were used. 
The probe laser frequency was scanned over the $5S_{1/2}-5P_{3/2}$ transition.
The pump laser frequency was fixed in the far wing of the atomic transition,
where the absorption length was much longer than the wavelength. 
The off-resonant optical excitation of rubidium atoms was incoherent due to
radiation trapping and interatomic excitation 
transfer~\cite{HolsteinPRA1947, Biberman1947, VASJQSRT2020}. 
As a result of comparison of experimental and calculated reflection spectra a
linear relation between self-broadening $\Gamma$ and the ground state population
$N_\text{g}$~\cite{VASPRL1996} is derived $\Gamma = K N_\text{g}$.

The observed excitation dependence is explained by the static mechanism of 
dipole-dipole interactions in the high-density atomic vapor~\cite{MukamelPRA1994}. 
Dipole-dipole interactions appeared only between the ground and excited atoms. 
Subsequent measurements~\cite{VASPRA1999, VASJPhysB2009} confirmed the derived 
linear relation for self-broadening ($\Gamma = K N_\text{g}$) in incoherently excited potassium and rubidium atomic vapors.

The optical resonance saturation of high-density rubidium vapors is investigated
in selective reflection experiments using 
one~\cite{VASJRLR2021, BobrovJETPLett2021, VASJQSRT2022} and two tunable 
lasers~\cite{VASNSRJQSRT2024}. 
Intense laser emission reduces both the width and magnitude of the selective reflection spectral profiles. 
Field-broadening effects are also discussed.

Interesting nonlinear phenomena may be observed in coherently excited resonant 
gas, when the Rabi oscillations dominate over relaxation 
processes~\cite{Demtroderbook2008, Boydbook2007}.
Observation of the coherent Autler--Townes effect is reported in the seminal 
publication~\cite{AutlerTownes1955}. 
The Autler--Townes components were observed in selective reflection from the 
high-density potassium vapor~\cite{VASATPRA2008}. 
The probe laser frequency was scanned over the $4S_{1/2} - 4P_{1/2}$ transition. 
The pump laser frequency was fixed at the $4S_{1/2} - 4P_{3/2}$ transition. 
The appearance of the Autler--Townes doublet was interpreted within the framework 
of "dressed atomic states" approach~\cite{CohenTannoudjiATbook1996,
CohenTannoudjibook1992}. 
The measured spectral splitting of the ground state $4S_{1/2}$ was defined by the
Rabi frequency.

It has been shown theoretically~\cite{KaiserPRA2019} that the "dressed dense
atomic gases" may be used to explore the many-body effects. 
The main goal of our current research is to identify experimental conditions
for the formation of the "dressed states" in the high-density atomic vapors. 
The optical preparation of new atomic levels can be described by a simple
semiclassical model~\cite{Demtroderbook2008}. 
In this framework, the optical coherent excitation of an ensemble of two-level 
atoms with an unperturbed transition frequency $\omega_{ab}$, a transition
dipole moment $D_{ab}$ and a homogeneous width $\gamma$ (FWHM) was considered. 
When the applied optical field of amplitude $E_0$ was tuned at the frequency $\omega_{ab}$, the atomic excitation rate was given by the
ratio of the optically saturated width $\gamma_{\text{sat}}/2$ and the resonance
Rabi frequency $\Omega_\text{R}$~\cite{Demtroderbook2008}:
\begin{equation}
    \Omega_\text{R}=\frac{D_{ab}E_0}{\hbar}.
    \label{eq1}
\end{equation}

For $\Omega_\text{R}>\gamma_{\text{sat}}/2$, Rabi oscillations 
lead to formation of sidebands at frequencies $\omega_{ab}\pm\Omega_\text{R}$. 
In the absorption spectrum, the symmetric resonances with a saturated width of
$\gamma_{\text{sat}}$ will appear at these frequencies. 
The frequency interval $\Delta\omega_\text{abs}$ between the absorption resonances 
is equal to $2\Omega_\text{R}$.
According to the "dressed atomic states" approach the strong optical field induces
new energy levels by dynamic Stark shifts~\cite{CohenTannoudjibook1992,
Boydbook2007}.

\begin{figure}[t]
    \centering
    \includegraphics{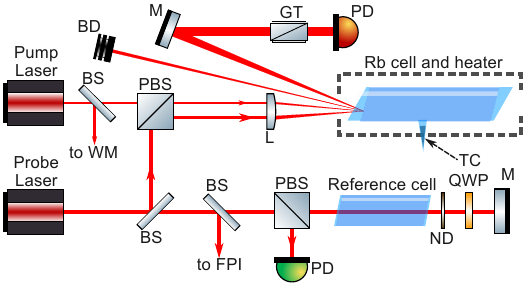}
    \caption{Optical layout of the setup. 
    Two orthogonally polarized laser beams from pump and probe lasers are 
    combined by polarizing beam splitter (PBS) and focused by lens (L) at 
    internal surface of the cell window. 
    The reflected probe beam is directed by a mirror (M) to a photodiode (PD)
    through Glan--Thompson polarizer (GT). 
    The frequency of the pump laser is measured by a wavemeter (WM,
    High-Finesse/Angstrom WS-U). 
    The frequency scan and single-mode regime of the probe laser are monitored
    by observing saturated absorption in a vapor cell and resonances of
    reference Fabry--Perot cavity (FPI). 
    The setup also includes the following optical elements: BS--beam splitter, 
    BD--beam dump, QWP--quarter-wave plate, ND--neutral-density filter.}
    \label{fig1}
\end{figure}

Analysis of selective-reflection spectra is more complicated than analysis of 
absorption spectra.
The spectral profile of unsaturated selective reflection can roughly be
approximated by a dispersive curve with a width of $\Gamma_0$~\cite{BoydPRL1991,
VASPRA1999}.
The spectral resolution in this case is limited by slowly decaying wings:
$R\propto(\omega-\omega_{ab})^{-1}$.
The derivative $\text{d}R/\text{d}\omega$ gives a bell-shaped spectral profile
(absorption-like resonance) with rapidly decreasing wings 
$\text{d}R/\text{d}\omega\propto(\omega-\omega_{ab})^{-2}$. 
Using this technique, the spectral resolution in the selective reflection
experiments can be substantially improved~\cite{VASATPRA2008}.
The resulting spectral curves can be analyzed by comparison with the
theoretical absorption spectra~\cite{Demtroderbook2008}.

In the presented study the derivative $\text{d}R/\text{d}\nu$ ($\nu = \omega/2\pi$) of the selective reflection coefficient $R$ was analyzed at several atomic densities and pump laser beam intensities. 
We take the derivative numerically from the spectral dependence of $R$.

\begin{figure}[t]
    \includegraphics[width=\linewidth]{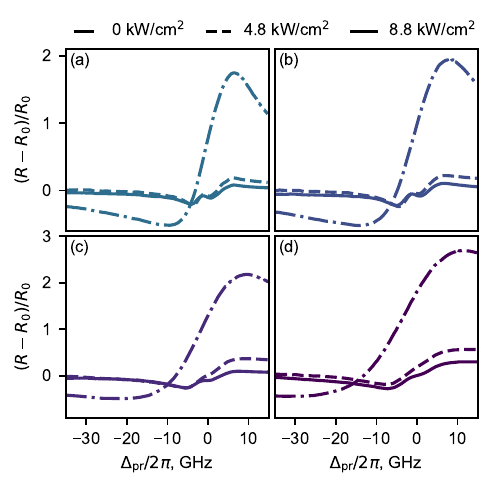}
    \caption{Spectral dependence of normalized reflection coefficient 
    $(R - R_0)/R_0$ for the probe beam, measured at four number densities: 
    (a)~$N_1 = 1.2 \times 10^{17}$~cm$^{-3}$ ($\Gamma_0/2\pi = 13.2$~GHz), 
    (b)~$N_2 =1.7 \times 10^{17}$~cm$^{-3}$ ($\Gamma_0/2\pi = 18.7$~GHz), 
    (c)~$N_3 = 2.5 \times 10^{17}$~cm$^{-3}$ ($\Gamma_0/2\pi = 27.5$~GHz), and 
    (d)~$N_4 = 3.6 \times 10^{17}$~cm$^{-3}$ ($\Gamma_0/2\pi = 39.6$~GHz).
    Zero frequency corresponds to the transition 
    $5S_{1/2} (F = 3)-5P_{3/2} (F' = 4)$ in $^{85}$Rb atoms.}
    \label{fig2}
\end{figure}

\begin{figure*}[ht]
    \centering
    \includegraphics[width=\linewidth]{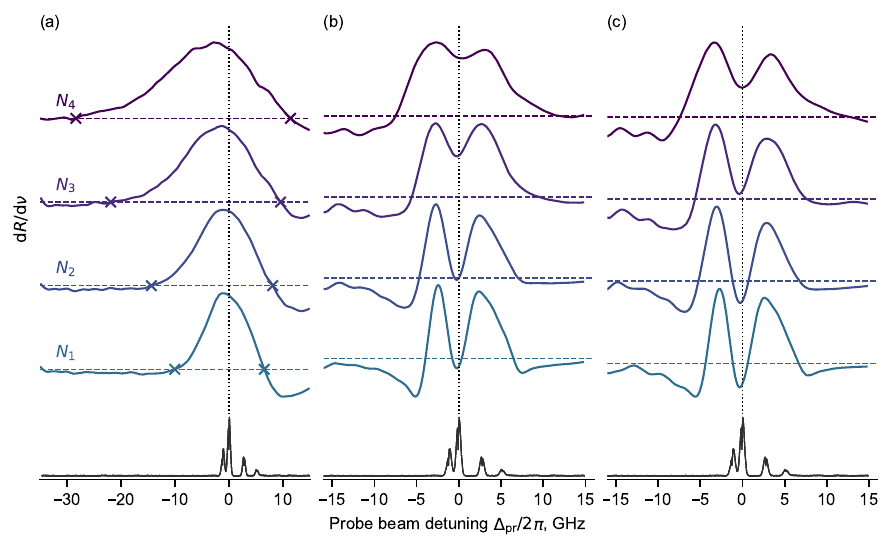}
    \caption{Frequency derivatives of the reflection coefficient 
    $\text{d}R/\text{d}\nu$ at four different number densities $N_1$, $N_2$, 
    $N_3$, $N_4$ and three selected values of the pump beam intensity
    (a)~$I_\text{pump}=0$~kW/cm$^2$,
    (b)~$I_\text{pump}=4.8$~kW/cm$^2$,
    (c)~$I_\text{pump}=8.8$~kW/cm$^2$.
    The bottom curves show saturated-absorption spectra of the probe beam for
    the reference rubidium cell. 
    The frequency scale in panel (a) is the same as in Fig.~\ref{fig2}.
    Zero frequency corresponds to the transition 
    $5S_{1/2} (F = 3)-5P_{3/2} (F' = 4)$ in $^{85}$Rb atoms.
    Horizontal dashed lines show the "zero" levels. 
    Vertical dotted lines indicate the location of the pump laser frequencies. 
    Marks (x) indicate crossing of the derivative curves with the "zero" levels (dashed lines).
    }
    \label{fig3}
\end{figure*}

\section{Experimental setup}

Nonlinear selective reflection from the window--vapor interface in the
high-temperature vapor cell with natural abundance of $^{85}$Rb and $^{87}$Rb
isotopes was studied at the transition $5S_{1/2}-5P_{3/2}$ (780~nm). 
This transition consists of two hyperfine (hfs) components from $^{85}$Rb and $^{87}$Rb
isotopes. 
A brief description of the used vapor cell is given in~\cite{VASselffoc2019}. 
The experiments are conducted at four atomic number densities $N_i$ in the range
$(1.2\text{--}3.6)\times10^{17}$~cm$^{-3}$ which defined by the temperatures of 
the cell from $367$ to $427^\circ$C.
Self-broadening $\Gamma_0=K N$ of the working transition is estimated by using 
the factor $K/2\pi = (1.1\pm0.17)\times10^{-16}$~GHz\,cm$^3$ from~\cite{AdamsJPhysB2011}.

The setup is practically the same as was used in Ref.~\cite{VASNSRJQSRT2024}. 
Spectroscopic measurements are performed using two tunable diode lasers: a 
low-power probe laser and a high-power pump laser. 
The probe laser frequency is scanned over all hfs components of
$5S_{1/2}-5P_{3/2}$ transition. 
To calibrate the probe laser frequency $\nu_\text{pr}$, we record the saturated
absorption spectrum of the reference rubidium vapor cell with natural abundance
of $^{85}$Rb and $^{87}$Rb isotopes at room temperature. 
In our measurements, the hyperfine (hfs) component $5S_{1/2} (F = 3) - 5P_{3/2} (F' = 4)$ in $^{85}$Rb atoms is chosen as the frequency reference $\nu_0$.
The pump laser frequency $\nu_\text{pump}$ is fixed at the reference $\nu_0$.
A schematic of the setup is shown in Fig.~\ref{fig1}.

The probe and pump beams with orthogonal linear polarization are combined by a polarizing beam splitter and then focused onto the window--vapor interface in the cell. 
The orthogonal polarizations are used to prevent possible formation of narrow resonances due to the coherent scattering of probe and pump optical fields by induced oscillations of atomic states population at the beat frequency~\cite{VASFWMPRA1997}. 
The contrast coherent resonances in a spectral dependence of the reflected probe beam were studied by using two beams with the same polarization~\cite{CASFWMQuEl2023}.
The selective reflection from the transparent dielectric--disordered gas interface is practically independent on a polarization of the incident optical beam near the normal angle of incidence.
The pump beam intensity can be varied from zero to 8.8~kW\,cm$^{-2}$. 
The probe beam intensity $I_\text{pr}$ is kept below 4~W\,cm$^{-2}$ to avoid a
distortion of the recorded curves which could be induced by the probe beam.
The probe beam is incident at 64~mrad. 
The angle of incidence for the pump beam is kept at 4~mrad. 
The probe beam reflected from the interface window/vapor is directed to a photodiode (PD).
Scattered radiation from the pump beam is additionally suppressed by a Glan-Thompson
polarizer (GT). 
The presence of the GT polarizer enhances contrast and does not change the measured selective reflection coefficient. 
The selective reflection coefficient was calibrated using the cell at room temperature and at large detuning of the pump beam frequency.
The reflection signal from the PD photodiode is recorded with a digital 
oscilloscope and a computer. 
The measured selective reflection coefficient $R$ is normalized by the 
nonresonant reflection coefficient $R_0\approx8.5$\% from the window/vacuum 
interface as $\delta R = (R - R_0)/R_0$. 
To ensure higher spectral resolution, we recorded the selective-reflection spectrum $R$ and averaged it using a digital oscilloscope to improve the signal-to-noise ratio.
The frequency derivative $\text{d}R/\text{d}\nu$ was then calculated numerically during data processing. 
The dispersive selective-reflection profile has slowly decaying wings, whereas its frequency derivative is bell-shaped with rapidly decreasing wings, thereby allowing closely spaced spectral features to be resolved more clearly.

\section{Experimental results and discussions}

The spectral dependencies of the normalized reflection coefficient $\delta R$,
recorded at different atomic number densities and pump intensities, are shown in
Fig.~\ref{fig2}.
The frequency scale is expressed as detuning 
$\Delta_\text{pr}/2\pi = (\nu_\text{pr}-\nu_0)$ of the probe laser from the hfs
component $5S_{1/2} (F = 3) - 5P_{3/2} (F' = 4)$.
The pump laser frequency $\nu_\text{pump}$ was fixed at the reference frequency
$\nu_0$.
The resonance optical saturation modifies the $\delta R$ spectral profiles, leading to
including splitting of the spectra.
The width of the unsaturated transition $5S_{1/2} - 5P_{3/2}$, which is a
combination of self-broadening and hfs-splitting, can be estimated as the
frequency interval between minimum and maximum of the $\delta R$ profiles in the 
absence of the pump beam~\cite{BoydPRL1991,VASPRA1999}.
More reliable results can be reached by measuring the frequency intervals 
$\Delta\nu_\text{SR}$ between "zeros" of the $\text{d}R/\text{d}\nu$ 
function~\cite{VASPRA1999}.
Calculated derivatives of the unsaturated reflection coefficient
$\text{d}R/\text{d}\nu$ are presented in Fig.~\ref{fig3}(a).
The marks (x) indicate crossing points of the unsaturated spectral dependence of derivative with the "zero" level (dashed line) in Fig.~\ref{fig3}(a).
According to~\cite{MukamelPRA1994} in the high-density atomic gas the dipole 
self-broadening can be represented as a combination of homogeneous and inhomogeneous widths. 
The hole-burning technique can help to estimate contributions of the homogeneous
and inhomogeneous parts to the linewidth. 
In the Figs.~\ref{fig3}(b) and~\ref{fig3}(c) the nonlinear spectra of frequency
derivative $\text{d}R/\text{d}\nu$ are presented.
The dips at the pump laser frequency split the spectra of 
$\text{d}R/\text{d}\nu$ into two resonances.
At densities $N_1$, $N_2$ and $N_3$ the resonances are asymmetric.
The spectral difference between resonances can be attributed to the influence
of the hfs components.
Such properties of the spectra indicate that the resonances at densities $N_1$,
$N_2$ and $N_3$ are inhomogeneously broadened.
Dips are the result of the optical saturation of the resonance 
transitions~\cite{Demtroderbook2008}.
The width of each dip is a superposition of the homogeneous part of
self-broadening and the field broadening~\cite{Demtroderbook2008}.

\begin{figure}[t]
    \includegraphics[width=\linewidth]{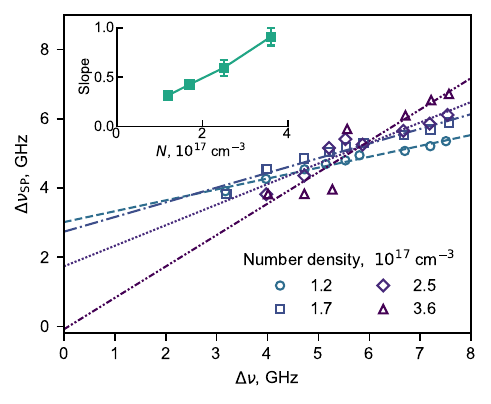}
    \caption{The values of measured splitting $\Delta\nu_{\text{SP}}$ are shown
    as data points.
    The lines are results of fit by a linear function
    $\Delta\nu_{\text{SP}}=a\Delta\nu+b$, where $a$--slope of the fitting 
    line, $b$--intercept of fitting line with the vertical axis.
    The inset shows the slope dependence on the atomic number density $N$.}
    \label{fig4}
\end{figure}

Curves in Figs.~\ref{fig3}(b), \ref{fig3}(c), which recorded at the maximal 
density $N_4$, demonstrate splitting of spectra of $\text{d}R/\text{d}\nu$ into
two symmetric resonances.
In our analysis, we consider the spectral width to be a more informative parameter than the amplitude. 
The double resonances with the same spectral widths can be considered as symmetric.
Which implies that the symmetric resonances are homogeneously broadened.
The spectral properties of the symmetric resonances can be explained either by the Rabi 
flopping oscillations in the simple semiclassical model~\cite{Demtroderbook2008}
or by the dynamic Stark shifts in the "dressed states" 
approach~\cite{CohenTannoudjibook1992,Boydbook2007}.
In both representations the frequency interval between symmetric resonances is
determined by the Rabi frequency.
Note that the resonance Rabi frequency $\Omega_\text{R}$ defined in Ref.~\cite{Demtroderbook2008} and the Rabi frequency $\Omega$ used in Refs.~\cite{CohenTannoudjibook1992,Boydbook2007} differ by a factor of two, such that 
$\Omega=2\Omega_\text{R}$.
For the working transition $5S_{1/2}-5P_{3/2}$ in rubidium atoms, the frequency
interval $\Delta\nu=2\Omega_\text{R}$ is related to the pump
intensity $I_\text{pump}$~\cite{VASJRLR2021} as follows:
\begin{equation}
    \Delta\nu~\text{[GHz]}=2.53\cdot(I_{\text{pump}}
    \text{[kW/cm}^{2}\text{]})^{1/2}.
    \label{eq2}
\end{equation}
The measurements of the splitting interval $\Delta\nu_\text{SP}$ between the
symmetric resonances confirm our suggestion, that new levels can be described as "atomic dressed states".
The experimental data and fitting lines are presented in Fig.~\ref{fig4}. 
As one can see on Fig.~\ref{fig3}(a) and Fig.~\ref{fig3}(b) for $N_1$ there is a small shift of the narrow symmetric resonances relative to the pump frequency. 
The frequency splitting is defined by the generalized Rabi frequency which depends on the resonance Rabi frequency and the frequency shift~\cite{Boydbook2007}. 
In our case the frequency shift induced contribution to the measured splitting is small relatively to experimental errors and can be neglected.

In the Table~\ref{tab} the fitted parameters $a_i$ and $b_i$ are presented for
the different atomic densities $N_i$.
\begin{table}[ht]
  \centering
  \begin{threeparttable}
    \caption{Fitted parameters for different atomic densities.}
    \label{tab}
    \begin{tabular}{l
                    S[table-format=1.1]
                    S[table-format=+1.2(2)]
                    S[table-format=+1.2(2)]}
      \toprule
      {$i$} & {$N_i$, $10^{17}$~cm$^{-3}$} & {Slope $a_i$} & 
      {Intercept $b_i$, GHz} \\
      \midrule
      1 & 1.2     & \num{0.31\pm0.02}       & \num{3.01\pm0.1}      \\
      2 & 1.7     & \num{0.42\pm0.03}       & \num{2.74\pm0.20}     \\
      3 & 2.5     & \num{0.59\pm0.08}       & \num{1.73\pm0.48}     \\
      4 & 3.6     & \num{0.93\pm0.16}       & \num{-0.20\pm0.95}    \\
      \bottomrule
    \end{tabular}
  \end{threeparttable}
\end{table}
The slopes of the fitted lines versus density are presented in the inset in 
Fig.~\ref{fig4}.
The theoretical slope, obtained by the expression (\ref{eq2}), is equal to
unity.
The difference between the "unity" and the slope of the fitted line for the
highest density $N_4$ is near 7\%.
This small difference supports our interpretation of the experimental results as
"dressed atomic states"~\cite{CohenTannoudjibook1992,Boydbook2007}.
In Table~\ref{tab} the small values of the slope $a$ at the densities $N_1$ and
$N_2$ reveal the inhomogeneous character of the line shapes.
The frequency intervals $\Delta\nu_{\text{SP}}$ in Fig.~\ref{fig4} are measured
between the maxima of the spectral dependences of the derivative $\text{d}R/\text{d}\nu$, separated by the optically saturated dips.
The corresponding intercepts $b_i$ at $N_1$ and $N_2$ in Table~\ref{tab} can be
associated with the homogeneous part of the widths.
The intermediate values of $a_3$ and $b_3$, measured at the density $N_3$, can be
explained by a transition from the inhomogeneous self-broadening to the
homogeneous self-broadening in this density region.
The transition from the inhomogeneous self-broadening (densities $N_1$, $N_2$ and $N_3$) to the homogeneous (density $N_4$) can be attributed to the quasi-molecular effects in this density region~\cite{VASZeemanPRA1997}. 
The quasi-molecular regime can appear in a density range where inter-atomic interactions are binary but the symmetry of the atomic wave function is changed due to the rapid succession of two-particle collisions. 
As a result, the atomic angular momenta could be modified and hfs components in atomic spectra could be reduced. 
We see an analogy between the observed reduction of hfs components in rubidium spectra (Fig.~\ref{fig3}) and the dependence of Zeeman splitting of $D$-lines in potassium vapor on the atomic density~\cite{VASZeemanPRA1997}.
It shall be noted that in our experiment a contribution of the quadrupole-quadrupole interaction from $5P-5P$ coupling to selective reflection spectra were not observed. 
Our data are consistent with the theoretical results of Ref.~\cite{MukamelPRA1994}, where higher multipole terms were neglected (e.g. the quadrupole-quadrupole interaction from the $5P-5P$ coupling). 
The calculations in Ref.~\cite{MukamelPRA1994} were performed for atomic density range in which an average distance between atoms is much larger than the Weisskopf radius.

\section{Conclusions}
This paper presents the results of an experimental study of nonlinear selective
reflection from high-density rubidium atomic vapor, where self-broadening of the
resonance transition dominates over the Doppler width.
In the experiment the pump-probe spectroscopic technique with two lasers are 
used. 
The reflection spectra of weak probe beam are recorded at several atomic 
densities in the range $(1.2 \text{--} 3.6)\times10^{17}$~cm$^{-3}$ and various 
pump beam intensities. 
As the density increased, the broadening mechanism transitioned from inhomogeneous to
homogeneous.
The strong pump field split the spectra into two resonances.
At the highest density these homogeneously broadened symmetric resonances are 
result from the Rabi flopping oscillations~\cite{Demtroderbook2008}.
The appearance of the symmetric resonances can be explained more strictly within the framework of "dressed atomic states" approach~\cite{CohenTannoudjibook1992,
Boydbook2007}.
At the highest density the experimental conditions are appropriated for research
of the dressed atomic states.
Now we can say that the main goal of our research is reached.
At lower densities the spectral profiles are inhomogeneously broadened.
Spectral profiles of the frequency derivative are separated by optically
saturated dips.
The width of such dips is a combination of the homogeneous part of
self-broadening and the intensity-dependent field broadening.
Additional studies of the transition from inhomogeneous to homogeneous
broadening may initiate a further development of the theory of interatomic 
interactions in a high-density atomic gas.

In nanocells the dressed atomic states can be investigated in absorption spectra of the dense atomic gases~\cite{SarkisyanPRL2012, SarkisyanPRL2018}.
The unique nonlinear properties of ultrathin layers of "dressed dense atomic
gases" might be useful for development of new quantum devices.
The "dressed states" spectroscopy may be applied to the study of many-body
interactions in ultracold nonideal plasmas with the resonance transitions in the
visible range~\cite{ZelenerJETPL2021, ZelenerPRL2024}.

\section*{CRediT authorship contribution statement}
\textbf{Vladimir Sautenkov}: Conceptualization, Methodology, Validation, 
                            Writing -- original draft.
\textbf{Sergey Saakyan}: Investigation, Software, Validation, Formal analysis,      
                        Writing -- review \& editing. 
\textbf{Andrei Bobrov}: Formal analysis, Methodology, Writing -- review \& editing. 
\textbf{Boris B. Zelener}: Supervision, Resources, Writing -- review \& editing,
                        Funding acquisition.

\section*{Declaration of Competing Interest}
The authors declare that they have no known competing financial interests or 
personal relationships that could have appeared to influence the work reported 
in this paper.

\section*{Data availability}
Data will be made available upon request.

\section*{Acknowledgements}
The research has been supported by the Ministry of Science and Higher Education 
of the Russian Federation (State Assignment No.\,075-00269-25-00).

\bibliographystyle{elsarticle-num}
\bibliography{refs}

\end{document}